\documentclass[aps, pre, notitlepage, superscriptaddress, twocolumn, 10pt, floatfix]{revtex4-2}
\usepackage[pdftex,plainpages=false,colorlinks=true,citecolor=blue,linkcolor=blue,urlcolor=black,filecolor=green,bookmarksopen=true]{hyperref}
\usepackage{amsmath}
\usepackage[caption = false]{subfig}
\usepackage{graphicx,epstopdf}
\usepackage[english]{babel}
\usepackage{blindtext}
\usepackage{lipsum}
\usepackage{amsfonts}
\usepackage{amssymb}
\usepackage{bbm}
\usepackage{enumerate}
\usepackage{color}
\usepackage{latexsym}
\usepackage{times,txfonts}
\usepackage{physics}

\DeclareSymbolFont{CMletters}{OML}{cmm}{m}{it}
\DeclareMathSymbol{J}{\mathalpha}{CMletters}{`J}
\DeclareMathSymbol{j}{\mathalpha}{CMletters}{`j}
\DeclareMathSymbol{U}{\mathalpha}{CMletters}{`U}

\begin{document}

\title{Quantum Maxwell's Demon Assisted by Non-Markovian Effects}

\author{Kasper Poulsen}
\email{poulsen@phys.au.dk}
\affiliation{Department of Physics and Astronomy, Aarhus University, Ny munkegade 120, 8000 Aarhus C, Denmark}

\author{Marco Majland}
\affiliation{Department of Physics and Astronomy, Aarhus University, Ny munkegade 120, 8000 Aarhus C, Denmark}

\author{Seth Lloyd}
\affiliation{Department of Mechanical Engineering, Massachusetts Institute of Technology, Cambridge, Massachusetts 02139, USA}

\author{Morten Kjaergaard}
\affiliation{The Niels Bohr Institute, Copenhagen University, Universitetsparken 5, 2100 Copenhagen, Denmark}

\author{Nikolaj T. Zinner}
\email{zinner@phys.au.dk}
\affiliation{Department of Physics and Astronomy, Aarhus University, Ny munkegade 120, 8000 Aarhus C, Denmark}
\affiliation{Aarhus Institute of Advanced Studies, Aarhus University, Høegh-Guldbergs Gade 6B, 8000 Aarhus C, Denmark}

\begin{abstract}

\vspace{0.2cm}

Maxwell's demon is the quintessential example of information control, which is necessary for designing quantum devices.
In thermodynamics, the demon is an intelligent being who utilizes the entropic nature of information to sort excitations between reservoirs, thus lowering the total entropy. 
So far, implementations of Maxwell's demon have largely been limited to Markovian baths.
In our work, we study the degree to which such a demon may be assisted by non-Markovian effects using a superconducting circuit platform. 
The setup is two baths connected by a demon-controlled qutrit interface, allowing the transfer of excitations only if the overall entropy of the two baths is lowered. 
The largest entropy reduction is achieved in a non-Markovian regime, and importantly, due to non-Markovian effects, the demon performance can be optimized through proper timing.
Our results demonstrate that non-Markovian effects can be exploited to boost the information transfer rate in quantum Maxwell demons.

\end{abstract}

\maketitle

\section{Introduction}
The thought experiment of Maxwell's demon has inspired countless discoveries since its conception by Maxwell more than 150 years ago \cite{leff2002Maxwell, knott1911life}. 
The original idea was to have two gases separated by a wall with a demon-controlled door. 
The demon lets particles through the door only if the overall entropy of the two gases is lowered \cite{bennett1987demons}. 
This seemingly defies the second law of thermodynamics, and the mechanism can only be explained by including the demon's information as entropy. 
Maxwell's demon, the Szilard engine, and variations thereof all rely on information as a resource to lower entropy and extract work \cite{szilard1929, PhysRevA.39.5378, PhysRevLett.106.070401, PhysRevLett.102.250602, Parrondo2015}.

Various versions of Maxwell's demon have been proposed theoretically \cite{PhysRevA.56.3374, PhysRevB.83.085428, PhysRevLett.110.040601, PhysRevLett.118.260603, Barato_2013, mandal2012work}, and the new found ability to control and manipulate quantum degrees of freedom has led to a wave of experimental realizations \cite{Cottet7561, PhysRevLett.113.030601, PhysRevLett.121.030604, PhysRevResearch.2.032025, Masuyama2018, PhysRevLett.115.260602}. 
Lately, other variants have also been proposed, e.g., a demon extracting heat using a gambling strategy \cite{PhysRevLett.126.080603} or a non-equilibrium system used as a demon to lower the entropy of a system \cite{PhysRevLett.123.216801}. 

So far, implementations of Maxwell's demon have been considered only for Markovian baths.  A Markovian bath is a bath whose evolution is memory free, i.e., the evolution of a system interacting with a Markovian bath depends only on the present state of the system \cite{breuer2002theory, PhysRevLett.101.150402, RevModPhys.89.015001}. Thus, all information flowing from the system to the bath is lost forever. By contrast, non-Markovian baths have memory effects which result in information backflow from the bath back into the system \cite{PhysRevA.97.032133, PhysRevA.94.010101, Pezzutto_2016, PhysRevLett.115.120403}. Previously, non-Markovian enhancement has been found for thermal engines \cite{PhysRevA.99.052106, Abah_2020}.

In our work, we study a direct analog to the original thought experiment consisting of two baths separated by a qutrit interface. Through the three steps of acquiring, using, and erasing information, a demon can autonomously transfer quanta of heat from one bath to the other only if the overall entropy of the two baths is lowered. This setup is a simple toy model for intuitively understanding the interplay between the demon memory and the decrease in entropy. 
We show that the maximum entropy decrease is achieved in the limit of weakly non-Markovian baths. Furthermore, the entropy decrease due to the demon can be assisted by the increased predictability and backflow of information of the non-Markovian baths. 
This is done by comparing the entropy reduction as a function of the demon's timing for both the Markovian and non-Markovian limits of the two baths. 
As we demonstrate below, this can be achieved in a small system realizable using several of the current quantum technology platforms.

\section{Setup}
\begin{figure}[t]
\centering
\includegraphics[width=1. \linewidth, angle=0]{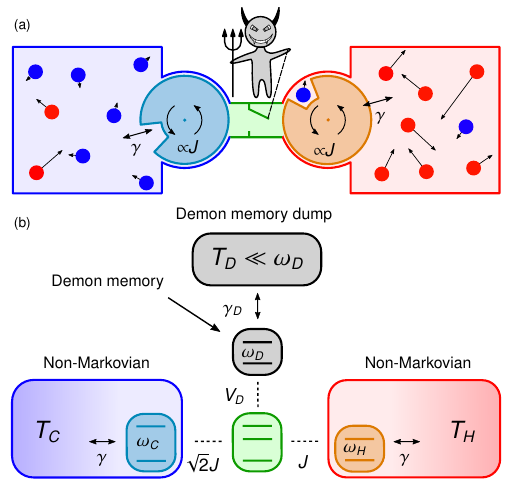}
\caption{(a) Classical analog of the demon setup where rotating wheels illustrate the added predictability of a non-Markovian bath.
(b) Illustration of the demon setup where two non-Markovian baths are connected by a qutrit. The cold non-Markovian bath consists of a qubit whose correlation functions decay due to the Markovian bath of temperature $T_C$ and likewise for the non-Markovian hot bath. A third qubit is demon memory, which can decay through interaction with the memory dump.
}
\label{figure1}
\end{figure}

The studied model is two non-Markovian baths connected by a qutrit, as seen in Fig.~\ref{figure1}. The non-Markovian baths are comprised of two parts: first, a Markovian bath of temperature $T_{C/H}$ and, second, a qubit with frequency $\omega_{C/H}$. A third qubit is used for demon memory. The Hamiltonian of the qutrit and the three qubits is given by
\begin{equation}
\begin{aligned}
\hat{H}_0 = \omega_C \Big[ \op{1_C}  +  \op{2_M}\Big] + \omega_H \Big[ \op{1_M} + \op{1_H}\Big]  \\
 + \omega_D \op{1_D}.
\end{aligned}
\end{equation}
The qutrit states are denoted $\ket{0_M}$, $\ket{1_M}$, and $\ket{2_M}$; the cold (hot) qubit states are denoted $\ket*{0_{C(H)}}$ and $\ket*{1_{C(H)}}$; and the demon-memory states are denoted $\ket{0_D}$ and $\ket{1_D}$. We are using units where $\hbar = k_B = 1$. The two qubits are coupled to the qutrit with strength $J$. If the qubit frequencies are picked such that $|\omega_C - \omega_H|, |\omega_C|, |\omega_H| \gg |J|$, the cold qubit can only couple the qutrit states $\ket{0_M}$ and $\ket{2_M}$, and the hot qubit can only couple the qutrit states $\ket{0_M}$ and $\ket{1_M}$. The full Hamiltonian becomes
\begin{equation}
\begin{aligned}
\hat{H} &= \hat{H}_0  + \sqrt{2} J\left( \hat{\sigma}_C^- |2_M\rangle \langle 0_M| + \hat{\sigma}_C^+ |0_M\rangle \langle 2_M|  \right) \label{hamilton} \\
&\hspace{2.5cm} + J\left( |0_M \rangle \langle 1_M| \hat{\sigma}_H^+ +  |1_M \rangle \langle 0_M| \hat{\sigma}_H^- \right),
\end{aligned}
\end{equation}
where $\hat{\sigma}_{C/H}^- = |0_{C/H}\rangle \langle 1_{C/H}|$ and $\hat{\sigma}_{C/H}^+ = |1_{C/H}\rangle \langle 0_{C/H}|$. The factor of $\sqrt{2}$ is due to the cold qubit interacting with the second excited state of the qutrit. The evolution of the system is described using the density matrix $\hat{\rho}$, through the Lindblad master equation \cite{Lindblad1976, breuer2002theory}
\begin{equation}
\frac{d \hat{\rho}}{d t} = -i [\hat{H}+\hat{V}_D (t), \hat{\rho}] + \mathcal{D}_C[\hat{\rho}] + \mathcal{D}_H[\hat{\rho}] + \mathcal{D}_D [\hat{\rho}](t). \label{me:1}
\end{equation}

\noindent The Markovian baths are modeled using the non-unitary parts
\begin{equation}
\begin{aligned}
\mathcal{D}_{C/H}[\hat{\rho}] &= \gamma (n_{C/H} + 1)  \left( \hat{\sigma}^-_{C/H} \hat{\rho} \hat{\sigma}^+_{C/H} - \frac{1}{2} \{ \hat{\sigma}^+_{C/H} \hat{\sigma}^-_{C/H}, \hat{\rho} \} \right)\\ &\hspace{1.5cm} + \gamma n_{C/H}   \left( \hat{\sigma}^+_{C/H} \hat{\rho} \hat{\sigma}^-_{C/H} -\frac{1}{2} \{ \hat{\sigma}^-_{C/H} \hat{\sigma}^+_{C/H}, \hat{\rho} \} \right), \nonumber \\ 
\mathcal{D}_D[\hat{\rho}](t) &= \gamma_D(t) \left( \hat{\sigma}^-_D \hat{\rho} \hat{\sigma}^+_D - \frac{1}{2} \{ \hat{\sigma}^+_D \hat{\sigma}^-_D, \hat{\rho} \} \right).
\end{aligned}
\end{equation}
The coupling strength between the Markovian baths and the cold and hot qubit is $\gamma$, the coupling of the demon memory to the memory dump is $\gamma_D(t)$, and the mean number of excitations in the bath mode of energy $\omega_C$ and $\omega_H$, respectively,~is
\begin{equation*}
n_{C} = \left( e^{\omega_C/T_{C}} -1 \right)^{-1} \quad \mathrm{and} \quad n_{H} = \left( e^{\omega_H/T_{H}} -1 \right)^{-1}.
\end{equation*}
To study the effects of non-Markovianity, we can keep the qutrit-bath coupling, $J$, constant while varying the rate of decay of the bath correlation functions through $\gamma$. The Markovian limit for the cold (hot) bath is $\gamma (n_{C(H)} + 1/2) \gg J$; see Appendix~\hyperref[appA]{A}.
If the system is left alone, i.e., $\hat{V}_D (t)= \gamma_D(t) = 0$, and the demon memory is reset to $\ket{0_D}$, the density matrix will eventually reach a unique steady state $\hat{\rho}_{\mathrm{ss}}$.
Unless otherwise stated, the parameters are suitably picked for superconducting circuits \citep{doi:10.1063/1.5089550} to be $J = 2\mathrm{MHz}$, $\omega_C = 7\mathrm{GHz}$, $\omega_H = 4\mathrm{GHz}$, $T_C = 4\mathrm{GHz} \simeq 31\mathrm{mK}$, and $T_H = 6\mathrm{GHz} \simeq 46\mathrm{mK}$. We also set $\gamma_D = 16\mathrm{MHz}$ when the demon memory is interacting with the memory dump and $\gamma_D = 0$ otherwise. However, everything is simulated using unitless variables and can be suitably rescaled. 

In summary, we study a cold non-Markovian bath interacting with the second excited state of the qutrit and a hot non-Markovian bath interacting with the first excited state of the qutrit. Excitations can thus be sorted from the cold to the hot bath by forcing the transition $\ket{2_M} \rightarrow \ket{1_M}$. 
This could be achieved through decay, which is equivalent to the transition being coupled to a bath at zero temperature. However, this would clearly result in heat flowing from the cold bath to this bath resulting in an entropy increase as expected. 

\section{Single shot} 
Instead, we wish to elucidate the interplay between entropy and information using a Maxwell's demon.
The demon memory is modeled by the qubit with frequency $\omega_D$. The demon operates in three steps:

\bigskip

\noindent \begin{minipage}{0.15\columnwidth}
Step 1.\\
\smallskip

Step 2.\\
\smallskip

Step 3.\\

\end{minipage}
\begin{minipage}{0.85\columnwidth}
Information on the qutrit is stored in the demon memory.
\smallskip

The information is used to transfer one excitation from the cold bath to the hot bath.
\smallskip

The demon memory is either reset or a clean memory slot is accessed.
\end{minipage}
\bigskip

\noindent For the qutrit in a general statistical mixture, the steps are
\begin{equation*}
\begin{aligned}
&\big(p_0 \op{0_M} + p_1 \op{1_M} + p_2 \op{2_M} \big) \op{0_D} \\
&\xrightarrow{\mathrm{step}\,1} p_0 \op{0_M 0_D} + p_1 \op{1_M 0_D} + p_2 \op{2_M 1_D} \\ &\xrightarrow{\mathrm{step}\,2} p_0 \op{0_M 0_D} + p_1 \op{1_M 0_D} + p_2 \op{1_M 1_D} \\ 
& \xrightarrow{\mathrm{step}\,3} \big(p_0 \op{0_M} + (p_1 + p_2) \op{1_M} \big) \op{0_D}
\end{aligned}
\end{equation*}
The first two steps constitute controlled NOT gates. These three steps will add energy to the system through work. Step 1 does average work $p_2 \omega_D$, and step 2 does average work $-p_2 (\omega_C - \omega_H)$. The work done through step 3 depends on how it is carried out. If the demon memory is reset through coupling to a cold bath, energy is subtracted through heat, and if a new demon memory is accessed, no heat or work is done. In either case, the total average work performed during the three steps is $p_2 (\omega_D + \omega_H-\omega_C)$. Thus work is performed to transfer heat similar to a refrigerator. However, looking at the special case $\omega_D = \omega_C-\omega_H$, we see that no work is done and the system does indeed implement a Maxwell's demon. For concreteness, we use a superconducting qubit platform to model an experimental implementation. 
\begin{figure}[t]
\centering
\includegraphics[width=1. \linewidth, angle=0]{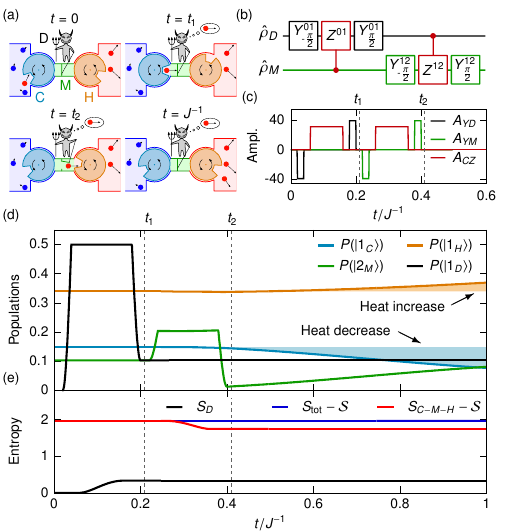}
\caption{Control, populations, and entropy for a single operation of the demon. The dashed lines separate step 1 ($0<t<t_1$), step 2 ($t_1<t<t_2$), and the subsequent free evolution ($t_2<t$). (a) Cartoon of the demon operation at different times. (b) Circuit diagram and (c) amplitudes, as seen in the control Hamiltonian \eqref{control}, for implementing steps 1 and 2 of the demon protocol.
(d) Populations for the excited states as a function of time starting from the system steady state at $t=0$. The orange and blue shadows show the difference between the population and the steady-state population for the hot and cold qubits. (e) Entropy of the baths and qutrit system $S_{C-M-H}$, the demon $S_D$, and the entire system $S_{tot}$ as a function of time. $\mathcal{S}$ is the constant entropy of the Markovian baths. For this simulation, $\gamma = 10^{-3}J$.}
\label{figure2}
\end{figure}
The CNOT gate can be implemented by supplementing the native controlled-phase gate \cite{DiCarlo2009} with single-qubit Y-gates; see  Fig. \ref{figure2}(b). The superconducting circuit control Hamiltonian relevant for this proposal can be written as
\begin{equation}
\begin{aligned}
\hat{V}_D(t) &= A_{YM}(t) \left(i |2_M \rangle \langle 1_M| e^{-i (\omega_C -\omega_H)t} - i |1_M \rangle \langle 2_M| e^{i (\omega_C -\omega_H) t} \right) \\ &\hspace{1cm}+ A_{YD}(t) \left( i |1_D \rangle \langle 0_D| e^{-i \omega_D t} - i |0_D \rangle \langle 1_D| e^{i \omega_D t} \right)\label{control} \\ &\hspace{1cm}+ A_{CZ}(t) \op{2_M 1_D}.
\end{aligned}
\end{equation}
\noindent The three amplitudes $A_{YM}$, $A_{YD}$, and $A_{CZ}$ define the demon protocol. These are picked such that the single-qubit Y-rotation gate time is $\tau_Y$ and the controlled-phase gate time is $\tau_{CZ}$. 
Unless otherwise stated, we set $\tau_Y = 0.02J^{-2} = 10\mathrm{ns}$ and $\tau_{CZ} = 0.1 J^{-1} = 50\mathrm{ns}$, which is achievable in superconducting circuits \citep{doi:10.1146/annurev-conmatphys-031119-050605, PhysRevX.11.021010}.
To show this process in action, the system is left alone for times $t<0$ such that the system reaches steady state $\hat{\rho}_{\mathrm{ss}}$, at $t=0$. Afterwards, step 1 and step 2 are implemented using the protocol shown in Fig.~\ref{figure2}(c).
The populations, $P(\ket{\alpha}) = \mathrm{tr}\Big\{ \op{\alpha} \hat{\rho} \Big\}$ for $\alpha~\in~\{ 1_C, 2_M, 1_H, 1_D \}$, are plotted for this process in Fig.~\ref{figure2}(d) with $\hat{\rho}=\hat{\rho}_{\mathrm{ss}}$ at $t=0$.  Here, $\mathrm{tr}\{\bullet\}$ denotes the trace over the entire Hilbert space.
From Fig. \ref{figure2}(d), we notice several things. 
After step 1, the demon-memory population reaches the value of the qutrit population, $P(\ket{1_D}) \sim P(\ket{2_M})$. 
After step 2, $P(\ket{2_M}) \sim 0$ and an excitation has been transferred from the cold to the hot bath, thus lowering the entropy of the baths and qutrit system. The transferred heat is also visible in the increase of $P(\ket{1_H})$ and the decrease of $P(\ket{1_C})$. The long-time behavior can be seen in Appendix~\hyperref[appB]{B}.
Without step 3, the demon memory is left in a statistical mixture giving $\ket{1_D}$ if an excitation was transferred and $\ket{0_D}$ otherwise. Moreover, the entropy of the baths and qutrit system, $S_{C-M-H}$, is lowered at the price of increasing the entropy of the demon memory, $S_D$. 
Both entropies together with the total entropy $S_{\mathrm{tot}}$ during the operation of the demon is plotted in Fig.~\ref{figure2}(e). The entropy of a system described by a density matrix $\hat{\rho}$ is defined by
\begin{equation}
S = - \sum_i \lambda_i \ln \lambda_i,
\end{equation}
where $\lambda_i$ are the eigenvalues of $\hat{\rho}$. 
The entropy $S_{C-M-H}$ does indeed decrease during the operation of the demon, the entropy of the demon increases, and the total entropy remains constant. Furthermore, the difference $S_{C-M-H} + S_{D} - S_{\mathrm{tot}}$ quantifies the mutual information between the qutrit-baths system and the demon memory. This mutual information is largest between steps 1 and 2, but it remains non-zero even after step 2. See Appendix~\hyperref[appC]{C} for a full discussion of the information flow.
Since the structure of the Markovian baths is unknown, their entropy is denoted $\mathcal{S}$, and the rate $\gamma \ll J$ is kept small enough that $\mathcal{S}$ can be assumed constant during the simulation. 
This implies that if we run the demon 
protocol once, as in Fig.~\ref{figure2}(d), all populations will eventually return to the steady state $\hat{\rho}_{\mathrm{ss}}$, 
as $T_C$ and $T_H$ are fixed. To calculate the change in temperature due to the exchange of energy quanta would require knowledge
of the heat capacity of the baths and depends on the concrete physical realizations, which are beyond the scope of the current discussion.
Without step 3, the demon protocol can only be run once, and the average number of excitations transferred will be less than 
\begin{equation}
\mathrm{tr}\Big\{ \op{2_M} \hat{\rho}_{\mathrm{ss}} \Big\} = \frac{e^{-\omega_C/T_C}}{1+e^{-\omega_H/T_H}+e^{-\omega_C/T_C}}.
\end{equation}
This does not exhibit any non-Markovian behavior since bath memory can not be seen through a single interaction.

\section{Non-Markovian effects} 
\begin{figure}[t]
\centering
\includegraphics[width=1. \linewidth, angle=0]{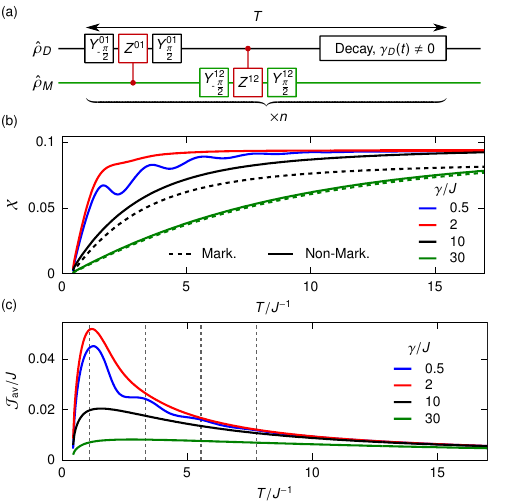}
\caption{Transferred heat as a function of demon timing due to non-Markovian effects. (a) Circuit diagram for implementing step 1, step 2, and step 3.
(b) Transferred excitations, $\mathcal{X}$, as a function of $T$ for different rates $\gamma$. This is plotted for both the full treatment (solid lines) and using a Markovian approximation on the cold and hot qubit (dashed lines). (c) Average excitation current, $\mathcal{J}_{\mathrm{av}}$, as a function of $T$ for different rates $\gamma$. 
}
\label{figure3}
\end{figure}

There are two ways to repeat the operation of the demon. First, the demon memory can be expanded. If the demon memory consists of $N$ qubits, the protocol can be repeated $N$ times. Second, information stored in the demon memory can be erased, allowing it to be reused. The demon memory is erased by letting it interact with the memory dump, i.e., $\gamma_D \neq 0$. 
We wish to study how the timing of the demon and the non-Markovian nature of the baths affect the transferred heat. Therefore, all three steps of the demon are repeated  without allowing the qubits to thermalize between cycles.
We let $T$ be the total time to perform all three steps. The three steps are repeated $n$ times such that when step 3 is finished, step 1 is performed once again. The new process is depicted in Fig. \ref{figure3}(a). 
To quantify the transport between the cold and hot baths, we define the  excitation current from the cold qubit to the qutrit as $\mathcal{J}_C = \mathrm{tr} \left\{ \hat{j}_C \hat{\rho} \right\}$, where $\hat{j}_C = -\sqrt{2}iJ\Big( \hat{\sigma}_C^- |2_M\rangle \langle 0_M |  - \hat{\sigma}_C^+ |0_M \rangle \langle 2_M| \Big)$, and the excitation current from the qutrit to the hot qubit as $\mathcal{J}_H = \mathrm{tr} \left\{ \hat{j}_H \hat{\rho} \right\}$, where $\hat{j}_H = -iJ\Big( |0_M\rangle \langle 1_M| \hat{\sigma}_H^+ - |1_M\rangle \langle 0_M| \hat{\sigma}_H^- \Big)$.
Since the Hamiltonian is time dependent, this will vary in time. To get a good measure of the number of transferred excitations, this is integrated over a single demon cycle,
\begin{equation}
\mathcal{X} = \lim_{n \rightarrow \infty} \int_{nT}^{(n+1)T} \mathcal{J}_C (t)\, dt = \lim_{n \rightarrow \infty} \int_{nT}^{(n+1)T} \mathcal{J}_H (t)\, dt.
\end{equation}
The integral above is the transferred excitations during the $n$th cycle of the demon. Even though the Hamiltonian is time dependent, the integral does converge for larger $n$; see Appendix~\hyperref[appD]{D}. From this, we also define the average excitation current, $\mathcal{J}_{\mathrm{av}} = \mathcal{X}/T$, driven by the demon. For large $T$, the system reaches the steady state $\hat{\rho}_\mathrm{ss}$ between each cycle and the transferred number of excitations is
\begin{equation*}
\lim_{T \rightarrow \infty} \mathcal{X} \leq \mathcal{X}_{ \mathrm{ss}}^{\mathrm{inst}} = \mathrm{tr}\Big\{ \op{2_M} \hat{\rho}_{\mathrm{ss}} \Big\} = \frac{e^{-\omega_C/T_C}}{1+e^{-\omega_H/T_H}+e^{-\omega_C/T_C}}.
\end{equation*}
$\chi_{\text{ss}}^{\text{inst}}$ is the transferred number of excitations only for instantaneous gates. In a realistic setting, the system is allowed to evolve during steps 1 and 2 resulting in less excitations transferred. Therefore, the actual number of transferred excitations, even in steady state, will be less than $\mathcal{X}_{ \mathrm{ss}}^{\mathrm{inst}}$.
$\mathcal{X}$ is plotted as a function of $T$ in Fig.~\ref{figure3}(b) for different values of $\gamma$. $T$ is varied through step 3 while the time to perform steps 1 and 2 is constant since it depends only on $\tau_Y$ and $\tau_{CZ}$. The dashed lines denote the case where the two qubits are traced away assumming that the states of the qubits are constant and thus Markovian, see Appendix~\hyperref[appA]{A}. 
Remarkably, the largest $\mathcal{X}$, and thus the largest entropy decrease, is achieved for non-Markovian baths, $\gamma = 2J$. The explanation for this is as follows: For small $\gamma$, the demon is limited by the small rate of excitation of the cold qubit. For large $\gamma$, the correlations between the cold qubit and the qutrit are suppressed resulting in a suppressed effective coupling between them that is similar to the quantum Zeno effect \cite{breuer2002theory}.
This explanation is further backed up in Appendix~\hyperref[appA]{A}. When the qubits start turning Markovian, $\gamma \geq 10J$, the full description predicts a larger $\mathcal{X}$ than the Markovian theory.
For $\gamma =30J$, the Markov approximation is valid and the results overlap. 
As $\gamma$ becomes small, $\mathcal{X}$ oscillates with $T$ due to non-Markovian effects or memory in the qubits. For $T$ sufficiently large, the system reaches the steady state between updates and the transferred excitations are the same for all $\gamma$. 

Another interesting quantity is the average current, $\mathcal{J}_{\mathrm{av}}$, which is plotted in Fig.~\ref{figure3}(c). 
Here the oscillations in $\mathcal{X}$ for smaller $\gamma$ are again clearly seen.
If the cold qubit is excited at $t = 0$, it will oscillate back and forth between the cold qubit and the qutrit. 
The excitation will be at the qutrit at times $t = \frac{\pi}{2 \sqrt{2}J} (1+2k)$, where $k \geq 0$ is a whole number. The first four of these times are drawn as dashed lines in Fig.~\ref{figure3}(c), which are close to the maxima in the oscillations. 
These oscillations are thus due to the non-Markovian nature of the cold bath.
The period of oscillation between the qutrit and hot bath is $\pi/J$. However, this period is not present in Figs.~\ref{figure3}(b) and \ref{figure3}(c), suggesting that the non-Markovian effects are predominantly due to the cold bath. This is further backed up in Appendix~\hyperref[appE]{E}. The largest entropy decrease is achieved with a combination of the larger coupling rate of Markovian baths and the increased predictability of non-Markovian baths.
This balance is met for $\gamma (n_C+1/2) \simeq J$. The precise value depends weakly on the hot bath temperature and two-qubit gate time; see Appendix~\hyperref[appF]{F}. However, the average current is robust towards changes in the coupling $\gamma$, and $\gamma (n_C+1/2) = J$ results in an average current only a few percent smaller than the maximum on most cases.
Since $\mathcal{X} > 0$ even in reverse bias, $T_C < T_H$, the system also implements a device of negative rectification, $\mathcal{R} = -\frac{\mathcal{J}_{\mathrm{av,f}}}{\mathcal{J}_{\mathrm{av,r}}} < 0$. Here, $\mathcal{J}_{\mathrm{av,f}}$ is the average current in forward bias, $T_C > T_H$, and $\mathcal{J}_{\mathrm{av,r}}$ is the average current in reverse bias, $T_C < T_H$. 
In order to resolve the non-Markovian dynamics and efficiently transfer excitation, we would expect to need gate times that are much shorter than the evolution of the system, $\tau_{CZ}, \tau_{Y} \ll J^{-1}$. In Appendix~\hyperref[appG]{G}, we find that the non-Markovian effects are seen for a wide range of gate times, $\tau_{CZ} \leq 0.4J^{-1}$, and cold bath temperatures. However, $\mathcal{X}$ only approaches the ideal, $\mathcal{X}_{\mathrm{ss}}^{\mathrm{inst.}}$, for $\tau_{CZ} \leq 0.2J^{-1} = 100\mathrm{ns}$, which is achievable for superconducting circuits.

\section{Conclusions} 
We have elucidated the interplay between entropy and information in the Maxwell's demon thought experiment using a simple cold bath, qutrit, and hot bath setup. Thus entropy can be decreased through three simple demon steps of acquiring, using, and deleting information. In deleting the information, the entropy of the memory dump is increased. Furthermore, we showed that the largest decrease in entropy is achieved on the border between Markovian and non-Markovian baths using a well-timed demon. 
This is due to a combination of two effects. First, the demon efficiency is limited by the effective coupling between the cold bath and the qutrit. This effective coupling is largest for balanced couplings, $\gamma (n_C+1/2) \simeq J$. Second, excitations oscillate back into the qutrit from the cold bath at certain times. By letting the demon operate at these times, the entropy decrease is boosted by the non-Markovian effects. Finally, we found that the demon can primarily be assisted by non-Markovian effects in the cold bath.

The setup can be implemented in superconducting circuits through four transmons: three concatenated to the lowest two levels and the fourth using the three lowest levels. All three qubits are coupled capacitively to the qutrit inducing hopping at resonance with strength $J$ as seen in the Hamiltonian \eqref{hamilton}. Single-qubit gates can be performed by capacitively coupling to a drive line, and a controlled-phase gate can be performed by using the avoided crossing of the higher excited levels. 

\begin{acknowledgments}
K.P., M.M., and N.T.Z. acknowledge funding from The Independent Research Fund Denmark DFF-FNU. M.K. acknowledges financial support through The Danish National Research Foundation and the Villum Foundation (Grant No. 37467) through a Villum Young Investigator grant.
\end{acknowledgments}

\section*{Appendix A: Markovian limit}
\label{appA}

We wish to calculate the Markovian limit of the two baths. To do this, the cold and hot qubits are assumed to have quickly decaying correlation functions such that they can be traced away.
First, we study the Markovian limit for just the cold qubit. The Hamiltonian of just the cold qubit is
\begin{align*}
\hat{H}_C = \omega_C \op{1_C} .
\end{align*}
In the case where this qubit is only weakly coupled to the rest of the system but strongly coupled to the heat bath, $J \ll \gamma$, the evolution of the density matrix of just the cold qubit $\hat{\rho}_C$ will predominantly be determined by the heat bath,
\begin{align*}
\frac{d\hat{\rho}_C}{dt} &= -i [\hat{H}_C, \hat{\rho}_C]  + \gamma n_C \left( \hat{\sigma}_C^+ \hat{\rho}_C \hat{\sigma}_C^- - \frac{1}{2} \{\hat{\sigma}_C^- \hat{\sigma}_C^+, \hat{\rho}_C\} \right) \\& \hspace{2.3cm} + \gamma (n_C + 1) \left( \hat{\sigma}_C^- \hat{\rho}_C \hat{\sigma}_C^+ - \frac{1}{2} \{\hat{\sigma}_C^+ \hat{\sigma}_C^-, \hat{\rho}_C\} \right),\\
n_C &= \left( e^{\omega_C/T_C} -1 \right)^{-1}.
\end{align*}
The state of the cold qubit will after sufficient time approach the thermal state,
\begin{align*}
\hat{\rho}_C(t \rightarrow \infty) &= \frac{e^{-\beta \hat{H}_C}}{\mathrm{tr} \{ e^{-\beta \hat{H}_C} \}} = (1-\lambda_C) \op{0}{0} + \lambda_C \op{1}, \\
\lambda_C &= \left(1 + e^{\omega_C/T_C}\right)^{-1}.
\end{align*}
In the Markovian limit, the cold qubit is assumed to remain in this state even for $J\neq 0$. The coherences between the qubit and the qutrit will decay exponentially in $\gamma$ and can therefore be neglected. In the Heisenberg picture, an operator $\hat{B}$ will evolve as
\begin{align*}
\frac{d}{dt}\hat{B}(t) &= i [\hat{H}_C, \hat{B}(t)]  + \gamma n_C \left( \hat{\sigma}_C^- \hat{B}(t) \hat{\sigma}_C^+ - \frac{1}{2} \{\hat{\sigma}_C^- \hat{\sigma}_C^+, \hat{\rho}_C\} \right) \\
&\hspace{2cm}+ \gamma (n_C + 1) \left( \hat{\sigma}_C^+ \hat{\rho}_C \hat{\sigma}_C^- - \frac{1}{2} \{\hat{\sigma}_C^+ \hat{\sigma}_C^-, \hat{\rho}_C\} \right).
\end{align*}
The Heisenberg picture is shown through the explicit time dependence. This can be solved for the ladder operators giving
\begin{align*}
\hat{\sigma}_C^-(t) &= \hat{\sigma}_C^- e^{-i\omega_C t - \gamma (n_C + 1/2) t},\\
\hat{\sigma}_C^+(t) &= \hat{\sigma}_C^+ e^{i\omega_C t - \gamma(n_C + 1/2) t}.
\end{align*}
With this the time correlation function, $\langle \hat{B}^\dag (t) \hat{B} \rangle$, for these two operators can be found to be
\begin{align}
\langle \hat{\sigma}_C^+(t) \hat{\sigma}_C^- \rangle &= \mathrm{tr}\{ \hat{\sigma}_C^+(t) \hat{\sigma}_C^- \hat{\rho}_C \} \nonumber\\
&= \lambda_C e^{i\omega_C t - \gamma(n_C + 1/2) t}, \label{corr} \\ 
\langle \hat{\sigma}_C^-(t) \hat{\sigma}_C^+ \rangle &= (1-\lambda_C) e^{-i\omega_C t - \gamma(n_C + 1/2) t}. \nonumber
\end{align}
The one-sided Fourier transforms are thus
\begin{align*}
\Gamma^+_C(\omega) &= \int_0^\infty dt\, e^{-i\omega t} \langle \hat{\sigma}_C^+(t) \hat{\sigma}_C^- \rangle \\
&= \lambda_C \int_0^\infty dt\, e^{i(\omega_C-\omega)t - \gamma(n_C+1/2) t} \\
&= \lambda_C \frac{i}{ \omega_C- \omega +i \gamma(n_C + 1/2)} \\
&= \lambda_C \frac{\gamma(n_C+1/2) + i(\omega_C- \omega)}{ (\omega_C- \omega)^2  + \gamma^2 (n_C + 1/2)^2}, \\
\Gamma^-_C(\omega) &= \int_0^\infty dt\, e^{-i\omega t} \langle \hat{\sigma}_C^-(t) \hat{\sigma}_C^+ \rangle\\
&= (1-\lambda_C) \frac{\gamma(n_C + 1/2) - i(\omega_C+ \omega)}{ (\omega_C + \omega)^2  + \gamma^2(n_C + 1/2)^2}.
\end{align*}
And thus
\begin{align*}
\gamma^+_C (\omega) &= \Gamma^+_C + \Gamma^{+*}_C
=\frac{\gamma \lambda_C (2n_C +1) }{ (\omega_C- \omega)^2  + \gamma^2 (n_C + 1/2)^2},\\
\gamma^-_C (\omega) &= \Gamma^-_C + \Gamma^{-*}_C
= \frac{\gamma (1-\lambda_C) (2n_C+1)}{ (\omega_C+ \omega)^2  + \gamma^2 (n_C+1/2)^2}.
\end{align*}
The same calculation can be carried out for the hot qubit,
\begin{align*}
\gamma^+_H (\omega) &=\frac{\gamma \lambda_H (2n_H +1) }{ (\omega_H - \omega)^2  + \gamma^2 (n_H + 1/2)^2},\\
\gamma^-_H (\omega) &= \frac{\gamma (1-\lambda_H) (2n_H+1)}{ (\omega_H + \omega)^2  + \gamma^2 (n_H+1/2)^2}.
\end{align*}
The interactions between the qutrit and two qubits are given by the terms
\begin{align*}
\hat{H}_{C-M} &= \sqrt{2} J\Big(\hat{\sigma}_C^+ |0_M \rangle \langle 2_M| + \hat{\sigma}_C^- |2_M \rangle \langle 0_M|\Big),\\
\hat{H}_{M-H} &= J\Big( |1_M\rangle \langle 0_M| \hat{\sigma}_H^-  + |0_M \rangle \langle 1_M| \hat{\sigma}_H^+ \Big).
\end{align*}
Treating the two qubits as environments and using the Redfield equation, after the Born-Markov and secular approximations, the master equation becomes
\begin{align*}
\frac{d \hat{\rho}}{d t} &= -i [\hat{H}_{0,m} + \hat{V}_D(t), \hat{\rho}] + \mathcal{D}_C[\hat{\rho}] +\mathcal{D}_H[\hat{\rho}] + \mathcal{D}_D (t)[\hat{\rho}],\\
\mathcal{D}_C[\hat{\rho}] &= \\
& \hspace{-0.1cm} 8 J^2 \frac{ 1-\lambda_C}{ \gamma (2n_C +1)} \left( |0_M \rangle \langle 2_M| \hat{\rho} |2_M \rangle \langle 0_M| - \frac{1}{2} \{ \op{2_M}, \rho \} \right)\\ &\hspace{-0.45cm} + 8 J^2 \frac{ \lambda_C}{ \gamma (2n_C +1)} \left( |2_M \rangle \langle 0_M| \hat{\rho} |0_M \rangle \langle 2_M| - \frac{1}{2} \{ \op{0_M}, \rho \} \right), \\ 
\mathcal{D}_H[\hat{\rho}] &= \\
& \hspace{-0.1cm} 4 J^2 \frac{ 1-\lambda_H}{ \gamma (2n_H +1)} \left( |0_M \rangle \langle 1_M| \hat{\rho} |1_M \rangle \langle 0_M| - \frac{1}{2} \{ \op{1_M}, \rho \} \right) \\
& \hspace{-0.45cm} + 4 J^2 \frac{ \lambda_H}{ \gamma (2n_H +1)} \left( |1_M\rangle \langle 0_M| \hat{\rho} |0_M \rangle \langle 1_M| - \frac{1}{2} \{ \op{0_M}, \rho \} \right), \\
\mathcal{D}_D[\hat{\rho}] &= \gamma_D(t) \left( \hat{\sigma}^-_D \hat{\rho} \hat{\sigma}^+_D - \frac{1}{2} \{ \hat{\sigma}^+_D \hat{\sigma}^-_D, \rho \} \right),\\
\hat{H}_{0,m} &= \omega_C \op{2_M} + \omega_H \op{1_M} + \omega_D \op{1_D}.
\end{align*}
Here, $\hat{V}_D(t)$ is the driving Hamiltonian. This approximation is valid when the correlation functions of the bath from Eq.~\eqref{corr} decay much faster than the dynamics of the system. Therefore, the inequality that needs to be fulfilled is 
\begin{align*}
\gamma (n_C + 1/2) \gg \sqrt{2} J \quad \mathrm{and} \quad \gamma (n_H + 1/2) \gg J
\end{align*}
for the cold and hot qubit, respectively. So the Markov approximation is not only valid for large $\gamma$, but also for large temperatures $T_C$ and $T_H$.

\begin{figure}[t]
\centering
\includegraphics[width=1. \linewidth, angle=0]{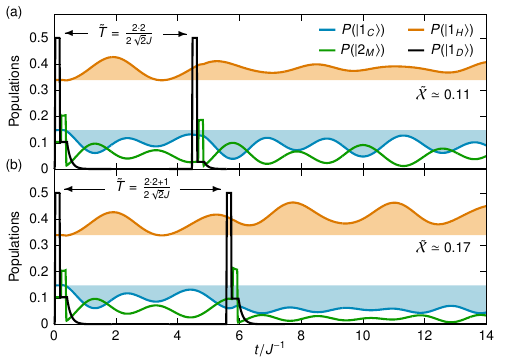}
\caption{Populations for the excited states as a function of time, starting from the system steady state at $t=0$. The figure is similar to Fig.~\ref{figure2}(d), however, the demon works twice here. The time between demon cycles is $\tilde{T}$ and the total number of transferred excitations is $\tilde{\mathcal{X}}$. For this simulation, $\gamma =10^{-3}J$.
}
\label{figure6}
\end{figure}

\section*{Appendix B: Double operation of the demon}
\label{appB}

Here we study the simplest operation where non-Markovian effects become important, that is, a double operation of the demon. The populations for this simulation can be seen in Fig.~\ref{figure6}. The protocol for the demon is the same as in Fig.~\ref{figure2}. However, in order for the demon to operate twice, the demon memory is allowed to decay between operations, as in Fig.~\ref{figure3}(a). 
The time between operations is denoted $\tilde{T}$. From the first demon operation and until the second demon operation, the populations are the same as in Fig.~\ref{figure2}(d). 
Here the oscillations between the qubits and the qutrit are clearly visible. In Fig.~\ref{figure6}(a), the second demon operation is at $t = \frac{2\cdot 2}{2 \sqrt{2}J}$, which is the time it takes one excitation at the cold qubit to oscillate to the qutrit and back twice. This results in a total of $\tilde{\mathcal{X}}\simeq 0.11$ transferred excitations. In Fig.~\ref{figure6}(b), the second demon operation is at $t = \frac{2\cdot 2+1}{2 \sqrt{2}J}$, which is the time it takes one excitation to perform 2.5 oscillations. This results in a total of $\tilde{\mathcal{X}}\simeq 0.17$ transferred excitations. This is the effect that is exploited in the full demon protocol. However, note that if $\gamma$ is made bigger, the oscillations become damped, and the plot will look different.

\section*{Appendix C: Information flow}
\label{appC}

\begin{figure}[t]
\centering
\includegraphics[width=1. \linewidth, angle=0]{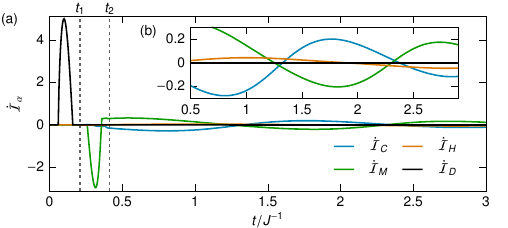}
\caption{(a), (b) Information rate $\dot{\mathcal{I}}_\alpha$ as a function of time for a single operation of the demon and $\gamma = 10^{-3}J$. }
\label{figure5}
\end{figure}

To quantify the information flow in the Maxwell's demon system, we define the multipartite mutual information,
\begin{align*}
\mathcal{I}_{C,M,H,D} = \sum_{\alpha \in \{ C, M, H, D \}} S_\alpha - S,
\end{align*}
where
\begin{align*}
S_\alpha &=- \text{tr} \{ \hat{\rho}_\alpha \ln \hat{\rho}_\alpha \}, \\
S &= -\text{tr} \{ \hat{\rho} \ln \hat{\rho} \}.
\end{align*}
The rate of change of this mutual information can be broken up into four contributions \cite{PhysRevLett.122.150603},
$\dot{\mathcal{I}}_{C,M,H,D} = \dot{\mathcal{I}}_C + \dot{\mathcal{I}}_M + \dot{\mathcal{I}}_H + \dot{\mathcal{I}}_D$, where
\begin{align*}
\dot{\mathcal{I}}_\alpha = -\text{tr} \{ \mathcal{L} [\hat{\rho}_\alpha] \ln \hat{\rho}_\alpha \} + \text{tr}\{ \mathcal{D}_\alpha [\hat{\rho}] \ln \hat{\rho} \}
\end{align*}
for $\alpha \in \{C, M, H, D\}$. $\hat{\rho}_\alpha$ is the density matrix for the $\alpha$ subsystem and $\mathcal{D}_M [\hat{\rho}] = 0$. The information rate is plotted in Figs.~\ref{figure5}(a) and \ref{figure5}(b) for a single operation of the demon. The three steps are clearly seen. For $0 < t < t_1$, information is gathered by the demon $\dot{\mathcal{I}}_D >0$. For $t_1 < t < t_2$, the information is used, thus lowering the information stored in the qutrit $\dot{\mathcal{I}}_M <0$. For $t_2 < t$, information oscillates between the qutrit and the hot and cold qubits. The oscillations are larger for the cold qubit since it is more non-Markovian than the hot qubit.

\section*{Appendix D: Convergence of the number of transferred excitations}
\label{appD}

\begin{figure}[t]
\centering
\includegraphics[width=1. \linewidth, angle=0]{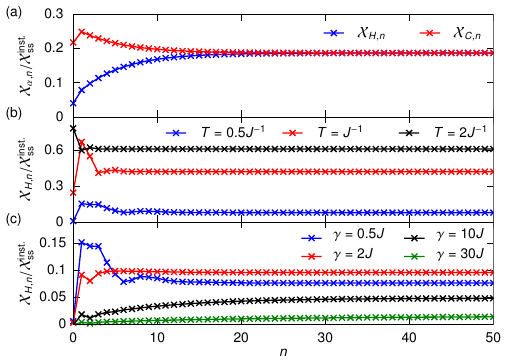}
\caption{Excitations transferred during the $n$th cycle as a function of $n$. (a) Transferred number of excitations between the cold bath and qutrit, $\mathcal{X}_{C,n}$, and between the qutrit and hot bath, $\mathcal{X}_{H,n}$, during the $n$th cycle as a function of $n$. Here, $\gamma = 10J$ and $T = J^{-1}$. (b) $\mathcal{X}_{H,n}$ as a function of $n$ for different $T$ and $\gamma = 0.5J$. (c) $\mathcal{X}_{H,n}$ as a function of $n$ for different $\gamma$ and $T = 0.5J^{-1}$. 
}
\label{figure4}
\end{figure}

In the main text, we looked at the limiting case where the demon protocol is used enough times such that the number of transferred excitations converge,
\begin{align*}
\mathcal{X} &= \lim_{n \rightarrow \infty} \int_{nT}^{(n+1)T} \mathcal{J}_C (t)\, dt = \lim_{n \rightarrow \infty} \int_{nT}^{(n+1)T} \mathcal{J}_H (t)\, dt.
\end{align*}
To check that this limit does indeed exist, we look instead at
\begin{align*}
\mathcal{X}_{C,n} &= \int_{nT}^{(n+1)T} \mathcal{J}_C (t)\, dt,\\
\mathcal{X}_{H,n} &= \int_{nT}^{(n+1)T} \mathcal{J}_H (t)\, dt.
\end{align*}
First, we plot $\mathcal{X}_{C,n}$ and $\mathcal{X}_{H,n}$ as a function of $n$ in Fig.~\ref{figure4}(a) for $\gamma = 10J$ and $T = J^{-1}$. It is seen that they both converge to the same value as expected. 
Next, $\mathcal{X}_{H,n}$ is plotted as a function of $n$ in Fig.~\ref{figure4}(b) for different values of $T$. $\mathcal{X}_{H,n}$ clearly converges for all values of $T$; however, convergence is slower for smaller $T$. 
Likewise, we plot $\mathcal{X}_{H,n}$ as a function of $n$ for different values of $\gamma$ in Fig.~\ref{figure4}(c). From this, we see that convergence is slower for $\gamma = 0.5J$ and $\gamma = 30J$. Due to these results, we choose to let $n\in [200,400]$ for $\gamma = 30J$ and $n\in [100,200]$ otherwise. The lower part of the interval is used for large $T$, while the upper part of the interval is used for smaller $T$. Furthermore, $\mathcal{X}$ is averaged over $10$ cycles.

\section*{Appendix E: Source of the non-Markovian effects}
\label{appE}

\begin{figure}[t]
\centering
\includegraphics[width=1. \linewidth, angle=0]{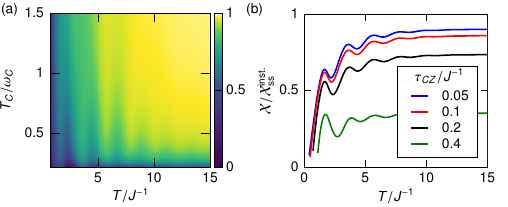}
\caption{(a) Transferred excitations $\mathcal{X}$ as a function of both cold bath temperature, $T_C$, and $T$ for $\gamma = 0.5J$ and $T_H = 1000J$. (b) Transferred excitations $\mathcal{X}$ as a function of $T$ for different controlled-phase gate times, $\tau_{CZ}$, and $\gamma = 0.5$. }
\label{figure8}
\end{figure}

To study which bath is the biggest source of the non-Markovian effects, $\mathcal{X}$ is plotted as a function of both $T$ and the cold bath temperature $T_C$ in Fig.~\ref{figure8}(c). 
For $T_C \ll \omega_C$, the cold bath is non-Markovian and the oscillations are observed. 
For $T_C > \omega_C$, the cold bath starts turning Markovian and the oscillations disappear. Therefore, the non-Markovian effects are mainly due to the cold bath. 
This is further supported by the fact that the oscillations were found to have a period of $\frac{\pi}{\sqrt{2}J}$ in the main article. As mentioned, this corresponds to excitations oscillating between the cold qubit and the qutrit. Excitations oscillating between the hot qubit and qutrit would have period $\frac{\pi}{J}$, which is not what we see.
Note that $T_H = 1000J \simeq 0.29\omega_C$ in Fig.~\ref{figure8}(c) such that $T_C > T_H$ in some cases. 
Since $\mathcal{X} > 0$ for both forward bias, $T_C > T_H$, and reverse bias, $T_C < T_H$, the system also implements a device of negative rectification, $\mathcal{R} = -\frac{\mathcal{J}_{\mathrm{av,f}}}{\mathcal{J}_{\mathrm{av,r}}} < 0$. Here, $\mathcal{J}_{\mathrm{av,f}}$ is the average current in forward bias, and $\mathcal{J}_{\mathrm{av,r}}$ is the average current in reverse bias. 

\section*{Appendix F: Optimal bath-qubit coupling rate}
\label{appF}

\begin{figure}[t!]
\centering
\includegraphics[width=1. \linewidth, angle=0]{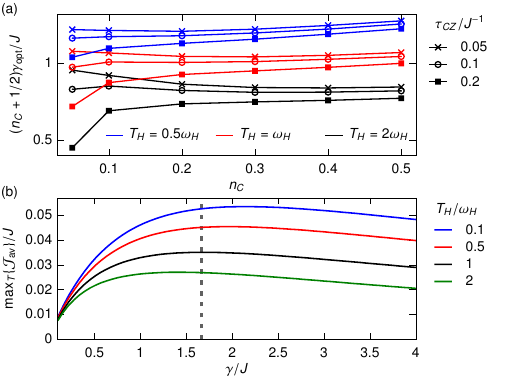}
\caption{(a) The product $(n_C +1/2)\gamma_{\text{opt}}$ as a function of $n_C$ for different values of $T_H$ and $\tau_{CZ}$. (b) The average current maximized over $T$, $\text{max}_T \{\mathcal{J}_{\text{av}}\}$, as a function of $\gamma$ for different hot bath temperatures and $n_C = 0.1$. The dashed line corresponds to $\gamma (n_C+1/2) = J$.}
\label{figure7}
\end{figure}

The optimal coupling rate is the value of $\gamma$ that allows for the largest average current induced by the demon, assuming that $T$ can be chosen freely. Therefore, we define the optimal coupling as
\begin{align}
\gamma_{\text{opt}} = \text{argmax}_\gamma \Big\{ \text{max}_T \{\mathcal{J}_\text{av} \}\Big\}.
\end{align}
From Eq.~\eqref{corr}, it is seen that the Markovianity of the cold bath is determined by the product $\gamma (n_C + 1/2)$. Therefore, the product $\gamma_{\text{opt}} (n_C + 1/2)$ as a function of $n_C = \left( e^{\omega_C/T_C} -1\right)^{-1}$ for different values of $T_H$ and $\tau_{CZ}$ is plotted in Fig.~\ref{figure7}(a).
Generally, the optimal coupling is seen to be around $\gamma_{\text{opt}} (n_C+1/2)\sim J$. However, the precise value depends on both the hot qubit temperature and the controlled-phase gate time. 
This is to be expected since the quality of the gates is influenced by both. For example, for larger $T_H$, the hot bath causes decoherence of the qutrit so a smaller $\gamma$ is preferred, whereas for small $T_H$, decoherence due to the hot bath is less important. In Fig.~\ref{figure7}(b), the average current maximized over the timing $T$ is plotted as a function of $\gamma$. Here is it seen that the precise value of $\gamma$ is not important. If $\gamma (n_C+1/2) = J$ is picked, the average current is within a few percent of the maximum that can be achieved in all four cases seen in Fig.~\ref{figure7}(b).

\section*{Appendix G: Effects of different gate times}
\label{appG}

In order to resolve the non-Markovian dynamics and effectively transfer excitation, we would expect to need gate times that are much shorter than the evolution of the system, $\tau_{CZ}, \tau_{Y} \ll J^{-1}$. 
Therefore, $\mathcal{X}$ is plotted in Fig.~\ref{figure8}(d) as a function of $T$ for different values of the controlled-phase gate time, $\tau_{CZ}$. 
We see that the non-Markovian dynamics is seen for all gate times. However, $\mathcal{X}$ only approaches the ideal, $\mathcal{X}_{\mathrm{ss}}^{\mathrm{inst}}$, for $\tau_{CZ} \leq 0.2J^{-1} = 100\mathrm{ns}$, which is achievable for superconducting circuits.

\bibliography{bibliography}

\end{document}